\documentclass[twocolumn]{autart}    

\usepackage{color}
\usepackage{pgfplots}
\usepackage{amsmath,amssymb}

\usepackage{tabularx}
\usepackage{booktabs}

\usepackage{tikz,pgfplots}
\pgfplotsset{compat=newest}
\pgfplotsset{plot coordinates/math parser=false}
\usepgfplotslibrary{fillbetween}
\usetikzlibrary{patterns}
\usepgfplotslibrary{colormaps}
\pgfplotsset{colormap={custom}{ rgb255=(64,64,64) rgb255=(205,205,205)}}   

\usepackage{nicematrix}
\usepackage{ifthen}
\usepackage{xspace}
\usepackage{stackengine}
\usepackage{xstring}
\usepackage{mathrsfs}  
\usepackage{scalerel}

\usepackage{algorithm}
\usepackage[noend]{algpseudocode}

\algrenewcommand\algorithmicindent{0.7em}%

\usepackage{textcomp}

\usepackage{todonotes}

\usepackage{subfig}

\usepackage[hidelinks]{hyperref}

\usepackage[para,online,flushleft]{threeparttable}

\usepackage{csquotes}                        
\newcommand\minusBig{-0.3ex}
\newcommand\symbolSubOpt{\dagger}
\newcommand\symbolLocalControlLaw{\kappa}

\newcommand\ubar[1]{\stackunder[0.25ex]{\hspace{0.15ex}$#1$}{\rule{0.75ex}{.1ex}}}
\newcommand*{\matrixBar}[1]{\stackunder[0.25ex]{\hspace{0.2ex}\ensuremath{#1}}{\rule{1.1ex}{.1ex}}}

\newcommand\vectorBracketLeft{(}
\newcommand\vectorBracketRight{)}

\newcommand*{\stateDim}{\ensuremath{p}\xspace}

\newcommand*{\inputDim}{\ensuremath{m}\xspace}

\newcommand*{\xV}{\stackunder[0.25ex]{\hspace{0.2ex}\ensuremath{x}}{\rule{0.75ex}{.1ex}} \xspace}

\newcommand*{\xo}{\stackunder[0.25ex]{\hspace{0.2ex}\ensuremath{x}}{\rule{0.75ex}{.1ex}}\ensuremath{_0}\xspace}

\newcommand*{\xf}{\stackunder[0.25ex]{\hspace{0.2ex}\ensuremath{x}}{\rule{0.75ex}{.1ex}}\ensuremath{_\mathrm{f}}\xspace}

\newcommand*{\xk}[1]{\stackunder[0.25ex]{\hspace{0.2ex}\ensuremath{x}}{\rule{0.75ex}{.1ex}}\ensuremath{(#1)}}

\newcommand*{\xplus}{\stackunder[0.25ex]{\hspace{0.2ex}\ensuremath{x}}{\rule{0.75ex}{.1ex}}\ensuremath{^+}\xspace}

\newcommand*{\xkDiff}[1]{\stackunder[0.25ex]{\hspace{0.2ex}\ensuremath{\check x}}{\rule{0.75ex}{.1ex}}\ensuremath{(#1)}}

\newcommand*{\xmu}[1]{\stackunder[0.25ex]{\hspace{0.2ex}\ensuremath{x}}{\rule{0.75ex}{.1ex}}\ensuremath{_\mu(#1)}}

\newcommand*{\xmuplus}{\stackunder[0.25ex]{\hspace{0.2ex}\ensuremath{x}}{\rule{0.75ex}{.1ex}}\ensuremath{^+_0}\xspace}

\newcommand*{\uV}{\stackunder[0.25ex]{\hspace{0.1ex}\ensuremath{u}}{\rule{0.75ex}{.1ex}}\xspace}

\newcommand*{\uf}{\stackunder[0.25ex]{\hspace{0.1ex}\ensuremath{u}}{\rule{0.75ex}{.1ex}}\ensuremath{_\mathrm{f}}\xspace}

\newcommand*{\uk}[1]{\stackunder[0.25ex]{\hspace{0.1ex}\ensuremath{u}}{\rule{0.75ex}{.1ex}}\ensuremath{(#1)}}

\newcommand*{\ukBlocked}[1]{\stackunder[0.25ex]{\hspace{0.1ex}\ensuremath{\check{u}}}{\rule{0.75ex}{.1ex}}\ensuremath{(#1)}}

\newcommand*{\ukf}[1]{\stackunder[0.25ex]{\hspace{0.1ex}\ensuremath{u}}{\rule{0.75ex}{.1ex}}\ensuremath{(#1)}}

\newcommand*{\ukfW}[2]{\stackunder[0.25ex]{\hspace{0.1ex}\ensuremath{\tilde{u}}}{\rule{0.75ex}{.1ex}}\ensuremath{(#1;#2)}}

\newcommand*{\ukfBlocked}[1]{\stackunder[0.25ex]{\hspace{0.1ex}\ensuremath{\check{u}}}{\rule{0.75ex}{.1ex}}\ensuremath{(#1)}}

\newcommand*{\ukfSubOpt}[2]{\stackunder[0.25ex]{\hspace{0.1ex}\ensuremath{u}}{\rule{0.75ex}{.1ex}}\ensuremath{^\symbolSubOpt(#1;#2)}}

\newcommand*{\ukfSubOptB}[2]{\stackunder[0.25ex]{\hspace{0.1ex}\ensuremath{u}}{\rule{0.75ex}{.1ex}}\ensuremath{^\symbolSubOpt\big(#1;#2\big)}}

\newcommand*{\ukfBlockedSubOpt}[2]{\stackunder[0.25ex]{\hspace{0.1ex}\ensuremath{\check{u}}}{\rule{0.75ex}{.1ex}}\ensuremath{^\symbolSubOpt(#1;#2)}}

\newcommand*{\refpair}{\ensuremath{(\xf,\uf)}\xspace}

\newcommand*{\refpairzero}{\ensuremath{(\ubar{0},\ubar{0})}\xspace}

\newcommand*{\extendedState}[1]{\ensuremath{\big\vectorBracketLeft #1,\uSW{#1} \big\vectorBracketRight}}

\newcommand*{\fv}{\stackunder[0.25ex]{\hspace{0.25ex}\ensuremath{f}}{\rule{0.75ex}{.1ex}}\xspace}

\newcommand*{\f}[3]{\stackunder[0.25ex]{\hspace{0.25ex}\ensuremath{f}}{\rule{0.75ex}{.1ex}}
	\IfEqCase{#3}{
		{s}{\ensuremath{(#1,#2)}}
		{b}{\ensuremath{\big(#1,#2\big)}}
		{B}{\ensuremath{\Big(\hspace{\minusBig}#1,#2\hspace{\minusBig}\Big)}}
	}
}  

\newcommand*{\openLoopSolutionf}[4]{\stackunder[0.25ex]{\hspace{0.2ex}\ensuremath{\varphi}}{\rule{0.75ex}{.1ex}}
	\IfEqCase{#4}{
		{s}{\ensuremath{(#1;#2,#3)}}
		{b}{\ensuremath{\big(#1;#2,#3\big)}}
		{B}{\ensuremath{\Big(\hspace{\minusBig}#1;#2,#3\hspace{\minusBig}\Big)}}
	}
}

\newcommand*{\unblockFunctionDef}{\stackunder[0.25ex]{\hspace{0.0ex}\ensuremath{\Theta}}{\rule{1ex}{.1ex}}\xspace}

\newcommand*{\limitInputFunctionf}[2]{\stackunder[0.25ex]{\hspace{0.25ex}\ensuremath{\Gamma}}{\rule{0.75ex}{.1ex}}
	\IfEqCase{#2}{
		{s}{\ensuremath{(#1)}}
		{b}{\ensuremath{\big(#1\big)}}
		{B}{\ensuremath{\Big(\hspace{\minusBig}#1\hspace{\minusBig}\Big)}}
	}
} 

\newcommand*{\unblockFunctionf}[2]{\stackunder[0.25ex]{\hspace{0.0ex}\ensuremath{\Theta}}{\rule{1ex}{.1ex}}
	\IfEqCase{#2}{
		{s}{\ensuremath{(#1)}}
		{b}{\ensuremath{\big(#1\big)}}
		{B}{\ensuremath{\Big(\hspace{\minusBig}#1\hspace{\minusBig}\Big)}}
	}
} 

\newcommand*{\totalWarmStartFunction}[2]{\stackunder[0.25ex]{\hspace{0.1ex}\ensuremath{\Omega}}{\rule{1.0ex}{.1ex}}
	\IfEqCase{#2}{
		{s}{\ensuremath{(#1)}}
		{b}{\ensuremath{\big(#1\big)}}
		{B}{\ensuremath{\Big(\hspace{\minusBig}#1\hspace{\minusBig}\Big)}}
	}
}

\newcommand*{\warmStartFunctionDef}{\stackunder[0.25ex]{\hspace{0.1ex}\ensuremath{\Omega}}{\rule{1.0ex}{.1ex}}\ensuremath{_\mathrm{sta}}\xspace}

\newcommand*{\warmStartFunction}[2]{\stackunder[0.25ex]{\hspace{0.1ex}\ensuremath{\Omega}}{\rule{1.0ex}{.1ex}}
	\IfEqCase{#2}{
		{s}{_\mathrm{sta}\ensuremath{(#1)}}
		{b}{_\mathrm{sta}\ensuremath{\big(#1\big)}}
		{B}{_\mathrm{sta}\ensuremath{\Big(\hspace{\minusBig}#1\hspace{\minusBig}\Big)}}
	}
} 

\newcommand*{\localWarmStartFunctionDef}{\stackunder[0.25ex]{\hspace{0.1ex}\ensuremath{\Omega}}{\rule{1.0ex}{.1ex}}\ensuremath{_\kappa}\xspace}

\newcommand*{\localWarmStartFunction}[2]{\stackunder[0.25ex]{\hspace{0.1ex}\ensuremath{\Omega}}{\rule{1.0ex}{.1ex}}
	\IfEqCase{#2}{
		{s}{_\kappa\ensuremath{(#1)}}
		{b}{_\kappa\ensuremath{\big(#1\big)}}
		{B}{_\kappa\ensuremath{\Big(\hspace{\minusBig}#1\hspace{\minusBig}\Big)}}
	}
}

\newcommand*{\stageCostf}[3]{
	\IfEqCase{#3}{
		{s}{\ensuremath{\ell(#1,#2)}}
		{b}{\ensuremath{\ell\big(#1,#2\big)}}
		{B}{\ensuremath{\ell\Big(\hspace{\minusBig}#1,#2\hspace{\minusBig}\Big)}}
	}
}

\newcommand*{\terminalCost}{\ensuremath{F}\ensuremath{(\cdot)}}

\newcommand*{\terminalCostDef}{\ensuremath{F}\xspace}

\newcommand*{\terminalCostf}[2]{
		\IfEqCase{#2}{
		{s}{\ensuremath{F(#1)}}
		{b}{\ensuremath{F\big(#1\big)}}
		{B}{\ensuremath{F\Big(\hspace{\minusBig} #1\hspace{\minusBig}\Big)}}
	}
}

\newcommand*{\totalCost}{\ensuremath{{J}_N}\ensuremath{(\cdot)}}

\newcommand*{\totalCostf}[3]{
	\IfEqCase{#3}{
		{s}{\ensuremath{{J}_N(#1,#2)}}
		{b}{\ensuremath{{J}_N\big(#1,#2\big)}}
		{B}{\ensuremath{{J}_N\Big(\hspace{\minusBig} #1,#2\hspace{\minusBig}\Big)}}
	}
}

\newcommand*{\totalCostExtndedStatef}[2]{
	\IfEqCase{#2}{
		{s}{\ensuremath{{J}_N(#1)}}
		{b}{\ensuremath{{J}_N\big(#1\big)}}
		{B}{\ensuremath{{J}_N\Big(\hspace{\minusBig} #1\hspace{\minusBig}\Big)}}
	}
}

\newcommand*{\optimalCostf}[2]{
	\ifthenelse{\equal{#1}{\xmu{n}} \OR \equal{#1}{\xmu{n+1}} \OR \equal{#1}{\xmu{n-1}} \OR \equal{#1}{\xmu{n+2}}  \OR \equal{#1}{\xmu{0}} \OR \equal{#1}{\xk{k}} \OR \equal{#1}{\xmu{k+1}}}
	{\ensuremath{{V}_{#2}\big(#1\big)}}
	{\ensuremath{{V}_{#2}(#1)}}
}

\newcommand*{\optimalCostBlockingf}[2]{
	\ifthenelse{\equal{#1}{\xmu{n}} \OR \equal{#1}{\xmu{n+1}} \OR \equal{#1}{\xmu{n-1}} \OR \equal{#1}{\xmu{n+2}}  \OR \equal{#1}{\xmu{0}} \OR \equal{#1}{\xk{k}} \OR \equal{#1}{\xmu{k+1}}}
	{\ensuremath{V^\mathrm{b}_{#2} \big(#1\big)}}
	{\ensuremath{V^\mathrm{b}_{#2} (#1)}}
}

\newcommand*{\normE}[1]{\ensuremath{\| #1 \|}}

\newcommand*{\normEQ}[2]{\ensuremath{\| #1 \|^2_{{\ubar{\scriptstyle{#2}}}}}}

\newcommand*{\alphaL}{\ensuremath{\alpha_1}\ensuremath{(\cdot)}}

\newcommand*{\alphaLf}[2]{
	\IfEqCase{#2}{
		{s}{\ensuremath{\alpha_1(#1)}}
		{b}{\ensuremath{\alpha_1\big(#1\big)}}
		{B}{\ensuremath{\alpha_1\Big(\hspace{\minusBig}#1\hspace{\minusBig}\Big)}}
		{s-1}{\ensuremath{\alpha^{-1}_1(#1)}}
		{b-1}{\ensuremath{\alpha^{-1}_1\big(#1\big)}}
		{B-1}{\ensuremath{\alpha^{-1}_1\Big(\hspace{\minusBig}#1\hspace{\minusBig}\Big)}}
	}
}

\newcommand*{\alphaU}{\ensuremath{\alpha_2}\ensuremath{(\cdot)}}

\newcommand*{\alphaUf}[2]{
	\IfEqCase{#2}{
		{s}{\ensuremath{\alpha_2(#1)}}
		{b}{\ensuremath{\alpha_2\big(#1\big)}}
		{B}{\ensuremath{\alpha_2\Big(\hspace{\minusBig}#1\hspace{\minusBig}\Big)}}
		{s-1}{\ensuremath{\alpha^{-1}_2(#1)}}
		{b-1}{\ensuremath{\alpha^{-1}_2\big(#1\big)}}
		{B-1}{\ensuremath{\alpha^{-1}_2\Big(\hspace{\minusBig}#1\hspace{\minusBig}\Big)}}
	}
}

\newcommand*{\alphaTerminal}{\ensuremath{\alpha_\mathrm{f}}\ensuremath{(\cdot)}}

\newcommand*{\alphaTerminalf}[2]{
	\IfEqCase{#2}{
		{s}{\ensuremath{\alpha_\mathrm{f}(#1)}}
		{b}{\ensuremath{\alpha_\mathrm{f}\big(#1\big)}}
		{B}{\ensuremath{\alpha_\mathrm{f}\Big(\hspace{\minusBig}#1\hspace{\minusBig}\Big)}}
		{s-1}{\ensuremath{\alpha^{-1}_\mathrm{f}(#1)}}
		{b-1}{\ensuremath{\alpha^{-1}_\mathrm{f}\big(#1\big)}}
		{B-1}{\ensuremath{\alpha^{-1}_\mathrm{f}\Big(\hspace{\minusBig}#1\hspace{\minusBig}\Big)}}
	}
}

\newcommand*{\alphaStageCost}{\ensuremath{\alpha_\ell}\ensuremath{(\cdot)}}

\newcommand*{\alphaStageCostf}[1]{\ensuremath{\alpha_\ell(#1)}}

\newcommand*{\mUf}[1]{\stackunder[-0.15ex]{\hspace{0.0ex}\ensuremath{\mu}}{\rule{0.55ex}{.1ex}}
\ifthenelse{\equal{#1}{\xmu{n}} \OR \equal{#1}{\xmu{n+1}} \OR \equal{#1}{\xmuDisturbed{n-1}} \OR \equal{#1}{\xmuDisturbed{n}} \OR \equal{#1}{\xmuDisturbed{n+1}} \OR \equal{#1}{\xmu{n-1}} \OR \equal{#1}{\xmu{n+2}}  \OR \equal{#1}{\xmu{0}}}
	{\ensuremath{\big(#1\big)}}
    {\ensuremath{(#1)}}
}

\newcommand*{\localControlLaw}{ \stackunder[0.25ex]{\hspace{0.15ex}\ensuremath{\symbolLocalControlLaw}}{\rule{0.55ex}{.1ex}} \ensuremath{(\cdot)}}
\newcommand*{\localControlLawDef}{ \stackunder[0.25ex]{\hspace{0.15ex}\ensuremath{\symbolLocalControlLaw}}{\rule{0.55ex}{.1ex}} \xspace}

\newcommand*{\localControlLawf}[1]{\stackunder[0.25ex]{\hspace{0.15ex}\ensuremath{\symbolLocalControlLaw}}{\rule{0.75ex}{.1ex}} 
\ifthenelse{\equal{#1}{\xmu{n}} \OR \equal{#1}{\xmu{n+1}} \OR \equal{#1}{\xmu{n-1}} \OR \equal{#1}{\xmu{n+2}}  \OR \equal{#1}{\xmu{0}} \OR \equal{#1}{\xk{k}} \OR \equal{#1}{\xk{k_1}} \OR \equal{#1}{\xk{k_2}} \OR \equal{#1}{\xmu{k+1}} \OR \equal{#1}{\xkDiff{k}} \OR \equal{#1}{\xkDiff{n}}}
	{\ensuremath{\big(#1\big)}}
    {\ensuremath{(#1)}}
}

\newcommand*{\localControlLawOpenLoopf}[2]{\stackunder[0.25ex]{\hspace{0.15ex}\ensuremath{\symbolLocalControlLaw}}{\rule{0.75ex}{.1ex}} 
	\IfEqCase{#2}{
	{b}{\ensuremath{\big(#1\big)}}
	{s}{\ensuremath{(#1)}}
}
}

\newcommand*{\xS}{\stackunder[0.25ex]{\hspace{0.1ex}\ensuremath{\mathbf{x}}}{\rule{0.95ex}{.1ex}}\xspace}

\newcommand*{\uS}{\stackunder[0.25ex]{\hspace{-0.05ex}\ensuremath{\mathbf{u}}}{\rule{0.95ex}{.1ex}}\xspace}

\newcommand*{\uSBlocked}{\stackunder[0.25ex]{\hspace{-0.05ex}\ensuremath{\check{\mathbf{u}}}}{\rule{0.95ex}{.1ex}}\xspace}

\newcommand*{\uSOpt}[1]{\stackunder[0.25ex]{\hspace{-0.05ex}\ensuremath{\mathbf{u}}}{\rule{0.95ex}{.1ex}}
	\ifthenelse{\equal{#1}{\xmu{n}} \OR \equal{#1}{\xmu{n+1}} \OR \equal{#1}{\xmu{n-1}} \OR \equal{#1}{\xmu{n+2}}  \OR \equal{#1}{\xmu{0}} \OR \equal{#1}{\xk{k}} \OR \equal{#1}{\xmu{k+1}}}
	{\ensuremath{^*\big(#1\big)}}
	{\ensuremath{^*(#1)}}
}

\newcommand*{\uSBlockedOpt}[1]{\stackunder[0.25ex]{\hspace{-0.05ex}\ensuremath{\check{\mathbf{u}}}}{\rule{0.95ex}{.1ex}}
	\ifthenelse{\equal{#1}{\xmu{n}} \OR \equal{#1}{\xmu{n+1}} \OR \equal{#1}{\xmu{n-1}} \OR \equal{#1}{\xmu{n+2}}  \OR \equal{#1}{\xmu{0}} \OR \equal{#1}{\xk{k}} \OR \equal{#1}{\xmu{k+1}}}
	{\ensuremath{^*\big(#1\big)}}
	{\ensuremath{^*(#1)}}
}

\newcommand*{\uSSubOpt}[1]{\stackunder[0.25ex]{\hspace{-0.05ex}\ensuremath{\mathbf{u}}}{\rule{0.95ex}{.1ex}}
	\ifthenelse{\equal{#1}{\xmu{n}} \OR \equal{#1}{\xmu{n+1}} \OR \equal{#1}{\xmu{n-1}} \OR \equal{#1}{\xmu{n+2}}  \OR \equal{#1}{\xmu{0}} \OR \equal{#1}{\xk{k}} \OR \equal{#1}{\xmu{k+1}}}
	{\ensuremath{^{\symbolSubOpt}\big(#1\big)}}
	{\ensuremath{^{\symbolSubOpt}(#1)}}
}

\newcommand*{\uSBlockedSubOpt}[1]{\stackunder[0.25ex]{\hspace{-0.05ex}\ensuremath{\check{\mathbf{u}}}}{\rule{0.95ex}{.1ex}}
	\ifthenelse{\equal{#1}{\xmu{n}} \OR \equal{#1}{\xmu{n+1}} \OR \equal{#1}{\xmu{n-1}} \OR \equal{#1}{\xmu{n+2}}  \OR \equal{#1}{\xmu{0}} \OR \equal{#1}{\xk{k}} \OR \equal{#1}{\xmu{k+1}}}
	{\ensuremath{^{\symbolSubOpt}\big(#1\big)}}
	{\ensuremath{^{\symbolSubOpt}(#1)}}
}

\newcommand*{\uSW}[1]{\stackunder[0.25ex]{\hspace{-0.05ex}\ensuremath{\tilde{\mathbf{u}}}}{\rule{0.95ex}{.1ex}}
    \ifthenelse{\equal{#1}{\xmu{n}} \OR \equal{#1}{\xmu{n+1}} \OR \equal{#1}{\xmu{n-1}} \OR \equal{#1}{\xmu{n+2}}  \OR \equal{#1}{\xmu{0}} \OR \equal{#1}{\xk{k}} \OR \equal{#1}{\xmu{k+1}}}
	{\ensuremath{\big(#1\big)}}
	{\ensuremath{(#1)}}
}

\newcommand*{\uSWsimple}[1]{\stackunder[0.25ex]{\hspace{-0.05ex}\ensuremath{\tilde{\mathbf{u}}}}{\rule{0.95ex}{.1ex}}
	\ifthenelse{\equal{#1}{\xmu{n}} \OR \equal{#1}{\xmu{n+1}} \OR \equal{#1}{\xmu{n-1}} \OR \equal{#1}{\xmu{n+2}}  \OR \equal{#1}{\xmu{0}} \OR \equal{#1}{\xk{k}} \OR \equal{#1}{\xmu{k+1}}}
	{\ensuremath{_\mathrm{w}\big(#1\big)}}
	{\ensuremath{_\mathrm{w}(#1)}}
}

\newcommand*{\uSWlocal}[1]{\stackunder[0.25ex]{\hspace{-0.05ex}\ensuremath{\tilde{\mathbf{u}}}}{\rule{0.95ex}{.1ex}}
	\ifthenelse{\equal{#1}{\xmu{n}} \OR \equal{#1}{\xmu{n+1}} \OR \equal{#1}{\xmu{n-1}} \OR \equal{#1}{\xmu{n+2}}  \OR \equal{#1}{\xmu{0}} \OR \equal{#1}{\xk{k}} \OR \equal{#1}{\xmu{k+1}}}
	{\ensuremath{_{\symbolLocalControlLaw}\big(#1\big)}}
	{\ensuremath{_{\symbolLocalControlLaw}(#1)}}
}

\newcommand*{\uSlocal}[1]{\stackunder[0.25ex]{\hspace{-0.05ex}\ensuremath{{\mathbf{u}}}}{\rule{0.95ex}{.1ex}}
	\ifthenelse{\equal{#1}{\xmu{n}} \OR \equal{#1}{\xmu{n+1}} \OR \equal{#1}{\xmu{n-1}} \OR \equal{#1}{\xmu{n+2}}  \OR \equal{#1}{\xmu{0}} \OR \equal{#1}{\xk{k}} \OR \equal{#1}{\xmu{k+1}}}
	{\ensuremath{_{\symbolLocalControlLaw}\big(#1\big)}}
	{\ensuremath{_{\symbolLocalControlLaw}(#1)}}
}

\newcommand*{\finiteInputSetExtendedHorizon}[1]{
	\ifthenelse{\equal{#1}{\xmu{n}} \OR \equal{#1}{\xmu{n+1}} \OR \equal{#1}{\xmu{n-1}} \OR \equal{#1}{\xmu{n+2}} \OR \equal{#1}{\xmu{0}}}
	{\ensuremath{\mathcal{A}\big(#1\big)}}
	{\ensuremath{\mathcal{A}(#1)}}
}

\newcommand*{\constrainedStateSpace}{\ensuremath{\mathbb{X}}\xspace}

\newcommand*{\constrainedInputSpace}{\ensuremath{\mathbb{U}}\xspace}

\newcommand*{\terminalSet}{\ensuremath{\mathbb{X}_\mathrm{f}}\xspace}

\newcommand*{\feasibleStateSpace}{\ensuremath{\mathcal{X}_N}\xspace}

\newcommand*{\feasibleStateSpaceMB}{\ensuremath{\bar{\mathcal{X}}_M}\xspace}

\newcommand*{\terminalLevelSet}[1]{\ensuremath{\mathrm{lev}_{#1}F}}

\newcommand*{\admissibleInputSpace}[2]{
	\ifthenelse{\equal{#1}{\xmu{n}} \OR \equal{#1}{\xmu{n+1}} \OR \equal{#1}{\xmu{n-1}} \OR \equal{#1}{\xmu{n+2}}  \OR \equal{#1}{\xmu{0}}}
	{\ensuremath{\mathcal{U}_{#2}\big(#1\big)}}
	{\ensuremath{\mathcal{U}_{#2}(#1)}}
}

\newcommand*{\admissibleWarmStartSpace}[1]{
	\ifthenelse{\equal{#1}{\xmu{n}} \OR \equal{#1}{\xmu{n+1}} \OR \equal{#1}{\xmu{n-1}} \OR \equal{#1}{\xmu{n+2}}  \OR \equal{#1}{\xmu{0}}}
	{\ensuremath{\tilde{\mathcal{U}}_N \big(#1\big)}}
	{\ensuremath{\tilde{\mathcal{U}}_N (#1)}}	
}

\newcommand*{\admissibleInputSpaceSubOptimal}[2]{\ensuremath{{\mathcal{U}}^\symbolSubOpt_{#2}\big(#1,\uSW{#1}\big)}}

\newcommand*{\admissibleTotalSpace}{\ensuremath{ \mathcal{Z}_N}\xspace}

\definecolor{gray1}{gray}{0.75}%
\definecolor{gray2}{gray}{0.65}
\definecolor{gray3}{gray}{0.55}
\definecolor{gray4}{gray}{0.45}
\definecolor{gray5}{gray}{0.35}

\newlength{\figureheight} 
\newlength{\figurewidth}  

\begin{document}

\begin{frontmatter}


\title{Suboptimal nonlinear model predictive control with input move-blocking} 


\author[Paestum]{Artemi Makarow}\ead{artemi.makarow@tu-dortmund.de},    
\author[Paestum]{Christoph R\"osmann}\ead{christoph.roesmann@tu-dortmund.de},               
\author[Paestum]{Torsten Bertram}\ead{torsten.bertram@tu-dortmund.de}  

\address[Paestum]{Institute of Control Theory and Systems Engineering, TU Dortmund University, 44227 Dortmund, Germany}  


\begin{abstract}                          
This paper deals with the integration of input move-blocking into the framework of suboptimal model predictive control. The blocked input parameterization is explicitly considered as a source of suboptimality. A straightforward integration approach is to hold back a manually generated stabilizing fallback solution in some buffer for the case that the optimizer does not find a better input move-blocked solution. An extended approach superimposes the manually generated stabilizing warm-start by the move-blocked control sequence and enables a stepwise improvement of the control performance. 
In addition, this contribution provides a detailed review of the literature on input move-blocked model predictive control and combines important results with the findings of suboptimal model predictive control. A numerical example supports the theoretical results and shows the effectiveness of the proposed approach. 
\end{abstract}

\end{frontmatter}


\section{Introduction}
\label{sec:introduction}

Nonlinear model predictive control (MPC) is a powerful control concept for complex cross-domain applications. Conventional MPC solves an optimal control problem (OCP) at every closed-loop time instance, including user-defined objective functions and nonlinear input and state constraints~\cite{Morari1999,Mayne2000}. Due to the advances in the fields of direct transcription and numerical optimization, MPC is emerging as an alternative control approach for mechatronic systems with fast dynamics and continuous objective functions, see, for example,~\cite{Verschueren2018,Andersson2018}. However, there is a certain class of nonlinear systems for which conventional MPC offers more degrees of freedom in control than required to achieve high control performance. An over-parameterized problem imposes an additional computational overhead. In particular, systems with small to mid-sized input and state dimensions and simple box-constraints are suitable for input move-blocking MPC (MBMPC). This strategy reduces the parameter optimization complexity by merging consecutive control vectors on the prediction horizon such that they have the same values along each dimension. A single control vector then represents a control block of constant length~\cite{Maciejowski2002,Tondel2002}. MBMPC is widely used, especially in industrial applications~\cite{Qin2003}. For some applications, even a single degree of freedom in control can be sufficient to satisfy the demands placed on the closed-loop control performance, see, for example,~\cite{Hampson1995,Makarow2018}. However, a constant partitioning of the predicted control sequence into blocks causes loss of recursive feasibility and asymptotic stability of the origin.  Conventional MPC with a receding horizon and stabilizing terminal conditions inherits the control invariance property of a terminal set by appending the local control law to the shifted and truncated (by one step) control sequence of the previous closed-loop time instance~\cite{Mayne2000,Rawlings2020}. Asymptotic stability of the origin mainly follows from the monotonicity property of the optimal value function~\cite{Mayne2000,Rawlings2020}. Unfortunately, both properties do not hold with move-blocking.
Another research direction that is highly relevant for practical applications is suboptimal MPC as first presented in~\cite{Scokaert1999}. Suboptimal MPC explicitly considers the situation that the optimizer cannot find the global optimum. This situation may arise with a non-convex OCP formulation or with limited computation times. Important closed-loop properties then rest upon manually generated stabilizing warm-starts~\cite{Scokaert1999,Pannocchia2011,Allan2017,Rawlings2020}. Suboptimal MPC that relies on stabilizing terminal conditions offers inherent robustness margins~\cite{Pannocchia2011,Allan2017} and addresses systems with continuous- and discrete-valued inputs~\cite{Rawlings2017}.

The major contribution of this paper is the integration of input move-blocking into the suboptimal MPC framework presented in~\cite{Rawlings2020,Allan2017}. This integration ensures important closed-loop properties like recursive feasibility and asymptotic stability of the origin for MBMPC with stabilizing terminal conditions. The optimal solution to an OCP subject to input move-blocking can be thought of as a suboptimal solution to the same OCP if the constraints arising from input move-blocking are removed. Thus, with this approach, input move-blocking is another source of suboptimality. Chen et al.~\cite{Chen2020} also notice this relationship, however, do not elaborate on this connection. By introducing some buffer in which the manually generated warm-starts can be stored between two consecutive closed-loop steps, important closed-loop properties of suboptimal MPC hold for MBMPC without further adjustment. Building on stabilizing warm-starts, the optimization time is assumed to have an upper bound as the optimization algorithm can be early-terminated~\cite{Rawlings2017}. This paper further combines suboptimal MPC with the embedding technique of the previous solution presented in~\cite{Ong2014,Shekhar2015}. The combined formulation then allows for stepwise improvement of the stabilizing warm-starts. The authors of ~\cite{Graichen2010} examine the stability properties of suboptimal MPC for continuous-time systems with input and without terminal constraints. In contrast to~\cite{Graichen2010}, this work does not require a lower bound on the number of optimization iterations for closed-loop stabilization. In addition, this article provides a detailed literature review on online MBMPC and reveals that asymptotic stability with MBMPC for nonlinear systems is still an open problem, in particular, when low computation time is important. 

The paper is organized as follows. The next Section~\ref{sec:literatur_review} provides a literature review on MBMPC. Afterward, Section~\ref{sec:suboptimal_mpc} integrates input move-blocking into the basic formulation of suboptimal MPC. Section~\ref{sec:improving_warmstart} presents an extension to suboptimal MPC to efficiently account for move-blocking. Section~\ref{sec:numerical_evaluation} supports the theoretical results by a numerical evaluation of a benchmark system. The paper closes with a conclusion in Section~\ref{sec:conclusion}.


\section{Literature review: MBMPC}
\label{sec:literatur_review}

\begin{table*}[h] 	
	\centering
	\caption{Literature review on input move-blocked online MPC. NTI: Nonlinear time-invariant, LTI: Linear time-invariant, RTI: Real-time-iteration, $+$: Statement applies, $-$: Statement does not apply, $\circ$: No detailed statement is made by the authors.} 	
	\label{tab:review_mbmpc} 	
	\vspace{0.0cm} 
	\begin{center} 	
		\resizebox{2.1\columnwidth}{!}{
			\begin{threeparttable}	
				\begin{tabular}{c |c |c| c| c| c| c| c} 	
					\toprule 	
					Reference, & Benchmark &  Applicability to  & Recursive     & Asymptotic  & Optimal($*$)/ & Sparsity/Structure     & Realization  \\	
					year & system    & NTI systems  &  feasibility  &  stability  & Suboptimal($\symbolSubOpt$)  & exploitation    & effort\tnote{5}       \\
					\midrule	
					Cagienard et al.~\cite{Cagienard2007}, 2007 &  LTI  &  $-$  & $+$ & $+$ & $*$ & $\circ$  & $\bullet$ \\
					\midrule	
					Gondhalekar et al.~\cite{Gondhalekar2007,Gondhalekar2009}, 2007(9) &  LTI  &  $+$  & $+$ & $-$ & $*$ &$\circ$  & $\bullet \bullet \tnote{6}$ \\
					\midrule
					Longo et al.~\cite{Longo2011}, 2011 &  LTI  &  $+$\tnote{1}  & $+$ & $+$ & $*$ & $\circ$  & $\bullet\,\bullet $ \\
					\midrule
					Shekhar et al.~\cite{Shekhar2015}\tnote{7}, 2015 &  LTI  &  $+$  & $+$ & $\circ$\tnote{2} & $*$ & $-$  & $\bullet$ \\
					\midrule
					Chen et al.~\cite{Chen2020}, 2019/20 &  NTI  &  $+$  & $-$ & $-$ & $\symbolSubOpt$ (RTI) & $+$  & $\bullet\bullet\bullet$ \\
					\midrule
					Gonzalez et al.~\cite{Gonzalez2020}, 2020 &  NTI  &  $+$  & $+$\tnote{3} & $+$\tnote{3}&$\symbolSubOpt$ (RTI)& $+$  & \mbox{$\bullet \bullet \bullet $} \\
					\midrule
					Son et al.~\cite{Son2020}, 2020 &  LTI  &  $-$  & $+$ & $+$ & $*$ & $\circ$  & $\bullet $ \\
					\midrule
					\midrule 
					Proposed approach &  NTI  &  $+$  & $+$ & $+$ & $\symbolSubOpt\tnote{4}$ &  $+$& $\bullet$ \\
					\bottomrule 	
				\end{tabular} 
				\begin{tablenotes}
					\item[1] Requires a sampled-data formulation and a uniform move-blocking pattern; \item[2] Proof of a valid Lyapunov function is not provided; \item[3] Derived from a terminal equality constraint; \item[4] Formulation also includes optimal solution; \item[5] Subjective assessment of the authors of this paper; \item[6] This rating considers LTI systems; \item[7] The evaluation in this row excludes optimal blocking (Sec.~4~in~\cite{Shekhar2015}).
				\end{tablenotes}
			\end{threeparttable}	
		}
	\end{center} 	
\end{table*}

Cagienard et al.~\cite{Cagienard2007} introduce a time-varying blocking scheme in combination with offset input move-blocking. Shifting the blocking pattern preserves the previous offset control interventions, while the base sequence and stabilizing terminal conditions ensure both recursive feasibility and asymptotic stability. Since the base sequence is generated with the linear-quadratic regulator (LQR), closed-loop properties mainly hold for linear time-invariant (LTI) systems. The authors of~\cite{Shekhar2012} also apply time-dependent blocking matrices in the framework of a variable horizon MBMPC. In~\cite{Gondhalekar2007,Gondhalekar2009} the authors investigate strong feasibility (see~\cite{Kerrigan2000}) issues in MBMPC. Since with MBMPC strong recursive feasibility cannot be derived from a control invariant terminal set, Gondhalekar and Imura~\cite{Gondhalekar2007} introduce the definition of controlled invariant feasibility. At each closed-loop time instance, this definition requires the first predicted state to be a member of the controlled invariant feasible set, such that the next OCP is guaranteed to be feasible under the conventional implicit control law. The authors of~\cite{Longo2011} exploit the sampled-data system formulation with piecewise constant inputs on a uniform time grid, integrate stabilizing terminal conditions and use two different time resolutions. The different time grids and a uniform blocking pattern allow one to formulate a discrete time system that is not subject to input move-blocking, and to realize a parallelizable execution of alternative move-blocked control sequences.  
Chen et al.~\cite{Chen2020} efficiently integrate the combination of input move-blocking and multiple shooting into the framework of the real-time iteration (RTI) scheme (see~\cite{Diehl2005} for RTI). The authors first embed move-blocking into the linearization step of the sequential-quadratic-programming method and then introduce a tailored condensing algorithm (e.g.,~\cite{Frison2016}) that accounts for the reduced degrees of freedom in control. However, the proposed integration does not consider important closed-loop properties, such as recursive feasibility. Gonzales and Rossiter~\cite{Gonzalez2020} also focus on embedding move-blocking into the RTI scheme. The authors choose single shooting and propose an admissible shifting strategy of the move-blocking pattern. In contrast to~\cite{Cagienard2007}, the proposed shifting strategy keeps the dimensions of the blocking matrix constant, although the blocking pattern alters as the closed-loop system evolves. This time-varying blocking pattern ensures recursive feasibility and asymptotic stability resulting from a terminal equality constraint. However, it is well-known that both move-blocking and the terminal equality constraint reduce the feasible set and thus the region of attraction of the controller (see, e.g.,~\cite{Shekhar2015,Rawlings2020}). The authors of~\cite{Ong2014} and~\cite{Shekhar2015} present an elegant way to include the shifted and truncated control sequence of the previous time instance into the current OCP formulation. By introducing additional optimization parameters, Shekhar and Manzie~\cite{Shekhar2015} show that the optimizer can either resort to the previous solution and append the local control law or improve the old solution by adding a proper move-blocked offset control sequence without shifting the blocking pattern. Building on this embedding technique, the authors of~\cite{Shekhar2015} prove recursive feasibility for MBMPC. The formal proof of asymptotic stability for nonlinear systems based on a suitable Lyapunov function candidate is pending. In particular, the authors do not discuss the case in which input move-blocking might be active inside the control invariant terminal set.  
Son et al.~\cite{Son2020} extend this embedding strategy by introducing a formulation that additionally interpolates between the previous solution and an LQR base sequence. This extension improves closed-loop performance, especially for LTI systems. Because of the LQR base sequence, this approach is only limitedly applicable to nonlinear systems. However, the optimizer can disable input move-blocking inside the terminal region by only relying on the LQR base control sequence (similar to~\cite{Cagienard2007}).  

Table~\ref{tab:review_mbmpc} summarizes important references on online MBMPC that include discussions on closed-loop properties. The last column of Table~\ref{tab:review_mbmpc} provides a relative and subjective ranking when the particular approach is implemented from scratch and the required effort is compared to the effort of the other approaches. 

The present work deals with the stability analysis of online MBMPC with a receding horizon for discrete-time systems. Therefore, the literature review does not cover papers on MBMPC that split the optimization into an offline and an online part (e.g.,\,\cite{Tondel2002,Goebel2014}), apply a non-uniform time discretization (e.g.,\,\cite{Yu2016}), use alternative parameterizations (e.g.,\,\cite{Rossiter2008}), or implement a shrinking horizon formulation~\cite{Farooqi2020}.


\section{Suboptimal MPC with input move-blocking}
\label{sec:suboptimal_mpc}
The nomenclature and the basic formulations are inspired by~\cite{Rawlings2020,Gruene2017}. This section adopts the stability assumptions from~\cite{Mayne2000,Rawlings2020}. The formulation of suboptimal MPC relies on~\cite{Rawlings2020,Pannocchia2011,Allan2017}.

\subsection*{Basic formulation of suboptimal MPC}
The following difference equation describes a nonlinear discrete-time system: 
\begin{equation}
	\xk{k+1}=\f{\xk{k}}{\uk{k}}{b}, \xk{0}=\xo.
	\label{eq:nonlinear_system}
\end{equation}
Equation~\eqref{eq:nonlinear_system} assigns a combination of a state vector $\xk{k} \in X := \mathbb{R}^\stateDim$ and an input vector $\uk{k} \in U := \mathbb{R}^\inputDim$ with $k\in \mathbb{N}_0$ to the successor state vector $\xk{k+1}$. The state and input spaces are Euclidean spaces with the dimensions $\stateDim\in\mathbb{N}$ and $\inputDim\in\mathbb{N}$, respectively.
\begin{assum}[Continuity of transition map]
	The transition map $\fv: X \times U \mapsto X$ is continuous. For some steady state $\refpair$, the transition map satisfies $\f{\xf}{\uf}{s} = \xf$.
	\label{assum:continuity_of_transition_map}
\end{assum}
Without loss of generality, we set the steady state to the origin $\refpairzero$ for short mathematical exposition. The Euclidean norm $\normE{\cdot}$ and the weighted norm $\normEQ{\xV}{Q}:=\xV^\intercal\matrixBar{Q}\xV$ with $\matrixBar{Q}\in \mathbb{R}^{\stateDim \times \stateDim}$ serve as distance metrics. The state vector $\xo \in X$ is used for initialization. Controlling the nonlinear system~\eqref{eq:nonlinear_system} by the sequence $\uS:=\big \vectorBracketLeft \ukf{0},\ukf{1},...,\ukf{N-1}\big \vectorBracketRight \in U^N$ with $N \in \mathbb{N}$, results in the state trajectory $\xS:=\big\vectorBracketLeft \xk{0},\xk{1},...,\xk{N}\big\vectorBracketRight\in X^{N+1}$. 
The function $\openLoopSolutionf{k}{\xo}{\uS}{s}$ invokes the iterative execution of equation~\eqref{eq:nonlinear_system} and thus represents the open-loop state trajectory at different time points. Because of Assumption~\ref{assum:continuity_of_transition_map}, the map $\vectorBracketLeft\xo,\uS \vectorBracketRight \mapsto \openLoopSolutionf{k}{\xo}{\uS}{b}$ is continuous (\cite[Prop.\,2.1]{Rawlings2020}). The controlled system must adhere to state and input constraints such that state vectors are admissible if $\xk{k}\in \constrainedStateSpace \subseteq X$ and input vectors are admissible if $\uk{k}\in \constrainedInputSpace\subset U$.
\begin{assum}[State and input constraint sets]
	The state constraint set $\constrainedStateSpace$ is closed. The terminal set $\terminalSet \subseteq \constrainedStateSpace$ is compact and contains the reference state vector $\xf$ in its interior. Finally, the input constraint set~$\constrainedInputSpace$ is compact and contains the reference input vector $\uf$.
	\label{assum:constrained_sets}
\end{assum}
The set of all admissible control sequences is defined by: 
\begin{equation}
	\begin{split}
		\admissibleInputSpace{\xo}{N}:=\{\uS\in \constrainedInputSpace^N\,|&\, \openLoopSolutionf{k}{\xo}{\uS}{b}\in \constrainedStateSpace,\\
		&\forall\,k = 0,1,...,N-1, \\
		&\openLoopSolutionf{N}{\xo}{\uS}{b} \in \terminalSet\}.
	\end{split}
	\label{eq:admissible_input_space}
\end{equation} 
From the definition of the set of all admissible control sequences, the feasible state set results in: 
\begin{equation}
	\feasibleStateSpace:=\{ \xo \in \constrainedStateSpace ~\,|\,~\admissibleInputSpace{\xo}{N} \neq \emptyset\}.
	\label{eq:feasible_set}
\end{equation}
We consider the following finite horizon cost function:
\begin{equation}
	\totalCostf{\xo}{\uS}{s} :=\sum\limits_{k=0}^{N-1} \stageCostf{\openLoopSolutionf{k}{\xo}{\uS}{b}}{\ukf{k}}{b}+\terminalCostf{\openLoopSolutionf{N}{\xo}{\uS}{s}}{b} .
	\label{eq:total_cost}
\end{equation}
\begin{assum}[Continuity of cost functions]
	The stage cost function $\ell:X \times U \mapsto \mathbb{R}_0^+$ and the terminal cost function $\terminalCostDef :X \mapsto \mathbb{R}_0^+$ represent continuous maps. To stabilize the origin with the nonlinear system~\eqref{eq:nonlinear_system}, the individual cost functions satisfy $\stageCostf{\ubar{0}}{\ubar{0}}{s} = 0$ and $\terminalCostf{\ubar{0}}{s}=0$. 
	\label{assum:continuity_of_cost_functions}
\end{assum}

A function $\alpha:\mathbb{R}_0^+\mapsto \mathbb{R}_0^+$ is said to be of class~$\mathcal{K}$ if it is continuous, strictly increasing and zero at zero. If it is unbounded in addition, then it is a member of the class~$\mathcal{K}_\infty$. 

\begin{assum}[Comparison functions]
	There exists a function $\alphaStageCost\in\mathcal{K}_\infty$ such that  $\stageCostf{\xV}{\uV}{b}\geq \alphaStageCostf{\normE{\big\vectorBracketLeft\xV,\uV\big\vectorBracketRight}}$ holds for all $\xV\in\constrainedStateSpace$ and $\uV\in\constrainedInputSpace$. There exists a function $\alphaTerminal\in\mathcal{K}_\infty$ such that $\alphaTerminalf{\xV}{b}\geq \terminalCostf{\xV}{b}$ holds for all $\xV\in \terminalSet$.
	\label{assum:comparison_functions}
\end{assum}

For costs in quadratic form with positive definite weighting matrices, the comparison functions follow directly from the scaled versions of the individual cost functions.

Stabilizing terminal conditions further build on a local control Lyapunov function (CLF) and constrain the final predicted state to a control invariant terminal set $\terminalSet$ (\cite[Asm. A1-A4]{Mayne2000} and~\cite[Asm. 2.14]{Rawlings2020}).

\begin{assum}[Control invariant terminal set]
	The compact terminal set $\terminalSet$ is control invariant for system~\eqref{eq:nonlinear_system}. The local control law $\localControlLawDef:\terminalSet\mapsto \constrainedInputSpace$ ensures: 
	\begin{equation}
		\xplus:=\f{\xV}{\localControlLawf{\xV}}{b}\in \terminalSet \mathrm{~if~}\xV\in\terminalSet.
		\label{eq:controlled_forward_invariance_terminal_set}
	\end{equation}
	The local controller renders the origin asymptotically stable such that $\terminalCost$ represents a local CLF for all $\xV\in\terminalSet$:
	\begin{equation}
		\terminalCostf{\xplus}{b}-\terminalCostf{\xV}{b} \leq -\stageCostf{\xV}{\localControlLawf{\xV}}{b}.
		\label{eq:clf}
	\end{equation}  
	\label{assum:controlled_forward_invariant_terminal_set}
\end{assum}
The authors of~\cite[Sec.\,2.5.5]{Rawlings2020} and \cite{Michalska1993,Chen1998} show that such a terminal region exists if the terminal cost function~$\terminalCost$ and the local control law $\localControlLaw$ are derived from linear system theory. Here, the main idea is to linearize the nonlinear system at the origin and design a linear state-space controller satisfying the algebraic Lyapunov equation. 

First, let $\uSW{\xo}\in\admissibleInputSpace{\xo}{N}$ be some admissible warm-start control sequence.
According to~\cite[Eq.\,(5b)]{Pannocchia2011} and~\cite[Eq.\,(12)]{Allan2017}, we define the set of all control sequences that are better than the warm-start in terms of costs by: 
\begin{align}
	\admissibleInputSpaceSubOptimal{\xo}{N} :=\big\{&\uS\in\admissibleInputSpace{\xo}{N}\,|\,\nonumber \\
	&\totalCostf{\xo}{\uS}{s} \leq   \totalCostf{\xo}{\uSW{\xo}}{b}\big\}.
	\label{eq:admissible_input_space_suboptimal}
\end{align}
Suboptimal MPC rests upon the following OCP:
\begin{equation}
	\min_{\uS\, \in \,\admissibleInputSpaceSubOptimal{\xo}{N}} \totalCostf{\xo}{\uS}{s}.  
	\label{eq:suboptimal_ocp}
\end{equation}
The globally optimal solution to OCP~\eqref{eq:suboptimal_ocp} is denoted by $\uSOpt{\xo}\in \,\admissibleInputSpaceSubOptimal{\xo}{N}$. However, we assume that the optimizer can also provide an admissible but suboptimal solution denoted by $\uSSubOpt{\xo}=\big\vectorBracketLeft\ukfSubOpt{0}{\xo},\ukfSubOpt{1}{\xo},...,\ukfSubOpt{N-1}{\xo}\big\vectorBracketRight\in \,\admissibleInputSpaceSubOptimal{\xo}{N}$ with $\totalCostf{\xo}{\uSOpt{\xo}}{b}\leq \totalCostf{\xo}{\uSSubOpt{\xo}}{b}$. Similar to the implicit control law in conventional MPC, the first element of the suboptimal control sequence selected by the optimizer is applied for closed-loop control:
\begin{equation}
	\xmuplus:=\xmu{n+1}=\f{\xmu{n}}{\ukfSubOptB{0}{\xmu{n}}}{B}.
	\label{eq:closed_loop_system}
\end{equation}
At each closed-loop time instance $n\in\mathbb{N}_0$, we set $\xo=\xmu{n}:= (x_{\mu,1}(n),x_{\mu,2}(n),...,x_{\mu,\stateDim}(n))^\intercal$. In the nominal case, hence, in the absence of model mismatch and external disturbances, the successor state $\xmuplus:=\xmu{n+1}$ initializes the OCP at time instance $n+1$. Therefore, the subsequent state $\xmuplus = \f{\xo}{\ukfSubOpt{0}{\xo}}{b}$ at time instance $n+1$ is already known at time instance $n$. Since there is often no guarantee that the optimizer will find the globally optimal solution, the authors of~\cite{Pannocchia2011,Allan2017} resolve the issue of how to design warm-start control sequences $\uSW{\xo}$ that ensure recursive feasibility and asymptotic stability of the origin (exponential stability in~\cite{Pannocchia2011}) without additional optimization. Note that $\uSW{\xo}\in\admissibleInputSpaceSubOptimal{\xo}{N}$. Therefore, the warm-starts must be designed to serve as stabilizing base solutions. In~\cite[Eq.\,(5c)]{Pannocchia2011} and \cite[Eq.\,(11)]{Allan2017}, the set of all admissible warm-start control sequences is defined by: 
\begin{equation}
	\begin{split}
	\admissibleWarmStartSpace{\xo}:=\{\tilde{\uS}\in\admissibleInputSpace{\xo}{N}\,|\,&\totalCostf{\xo}{\tilde{\uS}}{s}\leq\terminalCostf{\xo}{s}, \\
	&\mathrm{~if~} \xo \in \terminalSet\}.
	\end{split}
	\label{eq:admissible_warm_start_space}
\end{equation}
The admissible warm-start set $\admissibleWarmStartSpace{\xo}$ ensures the property that when $\normE{\xo}\to\ubar{0}$ it also follows that $\normE{\uSSubOpt{\xo}}\to\ubar{0}$~\cite{Pannocchia2011,Allan2017}. This property is essential for closed-loop stabilization when the implicit control law is based on suboptimal solutions (see~\cite[Lem.\,16]{Pannocchia2011} and~\cite[Prop.\,10]{Allan2017}). Let $\warmStartFunctionDef:\constrainedStateSpace\times\constrainedInputSpace^N \mapsto \constrainedInputSpace^N$ be the operator that shifts and truncates a control sequence by one step and then appends the local control law such that $\warmStartFunction{\xo,\uS}{s} = \big \vectorBracketLeft \ukf{1},\ukf{2},...,\ukf{N-1},\localControlLawOpenLoopf{\openLoopSolutionf{N}{\xo}{\uS}{s}}{b}\big \vectorBracketRight$. Let $\localWarmStartFunctionDef:\terminalSet \mapsto \constrainedInputSpace^N$ be the operator that applies the local control law for $N$ times such that $\localWarmStartFunction{\xo}{s} = \big \vectorBracketLeft \localControlLawf{\xo}, \localControlLawOpenLoopf{\f{\xo}{\localControlLawf{\xo}}{b}}{b},...\big\vectorBracketRight$. Now, suboptimal MPC generates warm-start solutions according to the following scheme and based on the prediction $\xmuplus = \f{\xo}{\ukfSubOpt{0}{\xo}}{b}$~(\cite[Eq.\,(5)]{Pannocchia2011} and~\cite[Eq.\,(13)]{Allan2017}):
\begin{equation}
	\uSW{\xmuplus} :=\begin{cases}
		\localWarmStartFunction{\xmuplus}{s} & \mathrm{if~} \xmuplus \in \terminalSet \mathrm{~and~} \\
		& \totalCostf{\xmuplus}{\localWarmStartFunction{\xmuplus}{s}}{b} \leq \\
		& \hspace{-0.5cm}\totalCostf{\xmuplus}{\warmStartFunction{\xo,\uSSubOpt{\xo}}{b}}{b}, \\
		\warmStartFunction{\xo,\uSSubOpt{\xo}}{b}    & \mathrm{otherwise}.                               
	\end{cases}
	\label{eq:generate_warmstarts}
\end{equation}
The strategy for generating warm-starts in~\eqref{eq:generate_warmstarts} is abbreviated by $\uSW{\xmuplus}:=\totalWarmStartFunction{\xo,\uSSubOpt{\xo}}{b}$. The first case in~\eqref{eq:generate_warmstarts} addresses the set definition in~\eqref{eq:admissible_warm_start_space} since $\localWarmStartFunction{\xo}{s} \in \admissibleWarmStartSpace{\xo}$~\cite[Prop. 9]{Pannocchia2011},~\cite[Prop.\,8]{Allan2017}. Finally, the authors of~\cite{Pannocchia2011,Allan2017} introduce the extended state $\ubar{z}:=\extendedState{\xo}$ and the following difference inclusion~\cite[Eq.\,14]{Allan2017}:
\begin{align}
	\ubar{z}^+\in\matrixBar{H}(\ubar{z}):=\big\{ \extendedState{\xmuplus}\,|\,& \xmuplus = \f{\xo}{\ukfSubOpt{0}{\xo}}{b}, \nonumber \\
	& \hspace{-0.5cm}\uSW{\xmuplus}=\totalWarmStartFunction{\xo,\uSSubOpt{\xo}}{b},\nonumber \\
	&\hspace{-0.5cm} \uSSubOpt{\xo}\in\admissibleInputSpaceSubOptimal{\xo}{N}\big\}.
	\label{eq:difference_inclusion}
\end{align}
The set-valued map $\matrixBar{H}(\cdot)$ includes the closed-loop system~\eqref{eq:closed_loop_system}, whose evolution is subject to an uncertain selection process of suboptimal solutions by the optimizer. Pannocchia et al.~\cite{Pannocchia2011} and Allan et al.~\cite{Allan2017} (see also~\cite{Rawlings2020}) show that $\totalCost$ is a Lyapunov function in the positive invariant set $\admissibleTotalSpace:=\{\extendedState{\xo}\,|\, \xo \in\feasibleStateSpace \mathrm{~and~} \uSW{\xo}\in\admissibleWarmStartSpace{\xo} \}$ for the closed-loop system~\eqref{eq:difference_inclusion}. When applying the implicit control law, the following inequalities hold with $\alphaL,\alphaU,\alpha_3(\cdot) \in \mathcal{K}_\infty$~\cite[Thm. 14]{Allan2017}: 
\begin{equation}
	\begin{gathered}
		\alphaLf{\normE{\ubar{z}}}{s}\leq \totalCostExtndedStatef{\ubar{z}}{s}\leq \alphaUf{\normE{\ubar{z}}}{s}, \\
		\sup_{\ubar{z}^+\in\matrixBar{H}(\ubar{z})} \totalCostExtndedStatef{\ubar{z}^+}{s} \leq  \totalCostExtndedStatef{\ubar{z}}{s} - \alpha_3(\normE{\ubar{z}}).
	\end{gathered}
	\label{eq:Lyapunov_function}
\end{equation}

The existence of the lower bound $\alphaL$ follows from Assumption~\ref{assum:continuity_of_transition_map}, Assumptions \ref{assum:continuity_of_cost_functions}-\ref{assum:comparison_functions}, and the algebraic transformations in~\cite[Prop.\,22]{Allan2017}. The upper bound $\alphaU$ results from Assumptions~\ref{assum:continuity_of_transition_map}-\ref{assum:comparison_functions} and~\cite[Prop.\,14]{Rawlings2017Tech}. Stabilizing terminal conditions in Assumption~\ref{assum:controlled_forward_invariant_terminal_set} ensure that $\totalCostExtndedStatef{\ubar{z}^+}{s}\leq \totalCostf{\xo}{\uSSubOpt{\xo}}{b} - \stageCostf{\xo}{\ukfSubOpt{0}{\xo}}{b} \leq \totalCostf{\xo}{\uSSubOpt{\xo}}{b} - \alphaStageCostf{\normE{(\xo,\ukfSubOpt{0}{\xo})}}$. Since $\uSSubOpt{\xo} \in \admissibleInputSpaceSubOptimal{\xo}{N}$, it follows that $\totalCostf{\xo}{\uSSubOpt{\xo}}{b}\leq \totalCostf{\xo}{\uSW{\xo}}{b} = \totalCostExtndedStatef{\ubar{z}}{s}$~\cite[Proof Thm. 14]{Allan2017}. Finally Allan et al. proof that there exists a function $\alpha_3(\cdot)\in \mathcal{K}_\infty$ such that $\alpha_3(\normE{\extendedState{\xo}}) \leq \alphaStageCostf{\normE{(\xo,\ukfSubOpt{0}{\xo})}}$~\cite[Prop.\,10, Proof Thm.\,14]{Allan2017}. It is important to note that the latter step explicitly requires that $\uSW{\xo}\in\admissibleWarmStartSpace{\xo}$. Asymptotic stability follows from~\cite[Prop.\,13]{Allan2017}.

\subsection*{Integrating input move-blocking}
Let $\matrixBar{B} \in \mathbb{R}^{N\times M}$ be the blocking matrix that describes the reduction of degrees of freedom in control from $N$ to $M\leq N$~\cite{Tondel2002}. It only contains zeros and ones. A uniform input move-blocking pattern with $M=2$ and a horizon length of $N=4$ adopts the following structure: 
\begin{equation}
	\matrixBar{B} := \begin{pNiceMatrix}
		1 & 1 & 0 & 0  \\
		0 & 0 & 1 & 1  
	\end{pNiceMatrix}^\intercal.
	\label{eq:blocking_matrix_example}
\end{equation}
According to the Definition 3 in~\cite{Cagienard2007}, a blocking matrix is admissible if it contains exactly one element equal to one in each row. Each new block, which follows the previous one in terms of prediction time, is indented by one column. The Kronecker product allows to describe the relationship between the blocked and the unblocked control sequence for systems with multiple inputs~\cite{Cagienard2007}:
\begin{equation}
	\uS = \unblockFunctionf{\uSBlocked}{s}:=\big( \matrixBar{B} \otimes \matrixBar{I}_{\inputDim} \big)\, \uSBlocked.
\end{equation} 
$\matrixBar{I}_{\inputDim}$ represents the identity matrix of input dimension~$\inputDim$ and the reduced order sequence is defined by $\uSBlocked:=\big \vectorBracketLeft \ukfBlocked{0},\ukfBlocked{1},...,\ukfBlocked{M-1}\big \vectorBracketRight \in U^M$. The vector field $\unblockFunctionDef: U^M \mapsto U^N$ expands the blocked sequence back to full length. Setting $M=N$ disables move-blocking such that $\uS = \unblockFunctionf{\uSBlocked}{s}=\uSBlocked$ holds. The formulation of input move-blocking with $M<N$ seamlessly integrates into the previous derivations on suboptimal MPC. The OCP formulation needs to be modified as follows:
\begin{equation}
	\begin{gathered}
		\min_{{\uSBlocked}} \totalCostf{\xo}{\unblockFunctionf{\uSBlocked}{s}}{b}, \\ 
		\mathrm{subject~to~} \unblockFunctionf{\uSBlocked}{s}\in\admissibleInputSpaceSubOptimal{\xo}{N}.  
	\end{gathered}
	\label{eq:suboptimal_mb_ocp}
\end{equation}
A suboptimal solution to OCP~\eqref{eq:suboptimal_mb_ocp} is denoted by
$\uSBlockedSubOpt{\xo}=\big\vectorBracketLeft\ukfBlockedSubOpt{0}{\xo},\ukfBlockedSubOpt{1}{\xo},...,\ukfBlockedSubOpt{M-1}{\xo}\big\vectorBracketRight$. Since input move-blocking is introduced additionally after OCP~\eqref{eq:suboptimal_ocp}, there is no guarantee that there exists a feasible solution to OCP~\eqref{eq:suboptimal_mb_ocp}. The following minor extension addresses the cases in which OCP~\eqref{eq:suboptimal_mb_ocp} turns infeasible:
\begin{equation}
	\uSSubOpt{\xo}:=\begin{cases}
			\unblockFunctionf{\uSBlockedSubOpt{\xo}}{b} & \mathrm{if~OCP~\eqref{eq:suboptimal_mb_ocp}~is~feasible},\\
		\uSW{\xo}  & \mathrm{otherwise}.
	\end{cases}
	\label{eq:fallback_level}
\end{equation}
In the case of $N=M$ and the application of continuous optimization (e.g., Newton-type solvers), the optimization algorithm usually accepts only those optimization steps that improve the warm-start in terms of costs (assuming a feasible optimization start). With $M<N$ and $M$ constant over all time steps $n$, the optimizer cannot handle the structure of the warm-start. Therefore, the optimizer either finds a better solution than the warm-start, starting from another point in the parameter space than the warm-start (first case in~\eqref{eq:fallback_level}), or the optimization routine needs to fall back onto the warm-start (second case in~\eqref{eq:fallback_level}). The term fallback solution is more accurate than the term warm-start in the setting of input move-blocking. Therefore, the manually generated fallback solution~$\uSW{\xo}$ must be temporarily buffered between two consecutive closed-loop steps. 
Note that the feasible set $\feasibleStateSpace$, which also forms the set $\admissibleTotalSpace$, does not consider input move-blocking in its formulation. To access the results from suboptimal MPC, the following assumption is necessary.
\begin{assum}[Admissible initial solution]
	At time instance $n=0$, there exists an admissible extended state $\extendedState{\xo}\in\admissibleTotalSpace$.
	\label{assum:initial_extended_state}
\end{assum}
This assumption is similar to the \enquote{oracle} in~\cite{Bobiti2017}, generating the first admissible warm-start, which is then improved by sampling-based optimization. 
\begin{prop}[Properties of MBMPC]
	Suppose Assumptions~\ref{assum:continuity_of_transition_map}-\ref{assum:initial_extended_state} hold. Assume that the suboptimal control sequences are generated based on equation~\eqref{eq:fallback_level}. Then, the feasible set~$\admissibleTotalSpace$ is positive invariant and the origin asymptotically stable in $\admissibleTotalSpace$ for the closed-loop system~\eqref{eq:difference_inclusion}. 
	\label{prop:properties_of_MBMPC}
\end{prop}
\begin{pf} 
	The fallback level in~\eqref{eq:fallback_level} includes the warm-start $\uSW{\xo}\in\admissibleWarmStartSpace{\xo}$ at every closed-loop time instance $n\geq 0$. Assumption~\ref{assum:initial_extended_state} ensures that there exits an initial admissible warm-start $\uSW{\xo}\in\admissibleWarmStartSpace{\xo}$ for all $\xo\in\feasibleStateSpace$, which is not necessary subject to input move-blocking. Since $\uSSubOpt{\xo}=\uSW{\xo}\in\admissibleInputSpaceSubOptimal{\xo}{N}$,  $\xmuplus\in\feasibleStateSpace$ holds after applying the suboptimal control vector $\ukfSubOpt{0}{\xo}$ for closed-loop control. From equation~\eqref{eq:generate_warmstarts} it follows that $\uSW{\xmuplus}=\totalWarmStartFunction{\xo,\uSW{\xo}}{b}\in\admissibleWarmStartSpace{\xmuplus}$ such that $\ubar{z}^+\in\admissibleTotalSpace$ (see~\cite[Proof Thm.\,14]{Allan2017}). The rest of the Proof (asymptotic stability) follows from~\cite[Prop.\,13 and Proof of Thm.\,14]{Allan2017} and relies on the Assumptions~\ref{assum:continuity_of_transition_map}-\ref{assum:controlled_forward_invariant_terminal_set}. 
\end{pf}
The rigorous realization of Assumption~\ref{assum:initial_extended_state} might require $M=N$ at time instance $n=0$. However, in practice, the initial states can be simply restricted to the following feasible set:
\begin{equation}
	\feasibleStateSpaceMB:=\{ \xo \in \constrainedStateSpace ~\,|\,\exists \uSBlocked\mathrm{~such~that~}\unblockFunctionf{\uSBlocked}{s} \in \admissibleInputSpace{\xo}{N}\}, 	
	\label{eq:feasible_set_mb_init}
\end{equation}
with $\feasibleStateSpaceMB\subseteq \feasibleStateSpace$. If $\xmu{0}\in\feasibleStateSpaceMB$ and $\xmu{0}\not\in\terminalSet$, then $\uSW{\xmu{0}}:=\unblockFunctionf{\uSBlocked}{s}\in\admissibleWarmStartSpace{\xmu{0}}$. If $\xmu{0}\in\terminalSet$, then $\uSW{\xmu{0}}:=\localWarmStartFunction{\xmu{0}}{b}\in\admissibleWarmStartSpace{\xmu{0}}$ (see~\cite[Algo.~9]{Allan2017}). In both cases, $\extendedState{\xmu{0}}\in\admissibleTotalSpace$. 
\begin{rem}[Blocking inside terminal set]
	If input move-blocking is active inside the terminal set $\terminalSet$, there is no guarantee that   $\totalCostf{\xmuplus}{\warmStartFunction{\xo,\unblockFunctionf{\uSBlockedSubOpt{\xo}}{b}}{b}}{b} \leq \terminalCostf{\xmuplus}{s}\leq \alphaTerminalf{\normE{\xmuplus}}{s}$ holds with $\xmuplus\in\terminalSet$. Therefore, input move-blocking relies, in particular, on evaluating $\localWarmStartFunction{\xmuplus}{s}$ $($see~\eqref{eq:generate_warmstarts}$)$ as an alternative warm-start.
\end{rem}
\begin{rem}[Inherent robustness]
	Allan et al.~\cite{Allan2017} derive inherent robustness properties for suboptimal MPC with $\constrainedStateSpace:=X$ (softened state constraints) and $\terminalSet:=\terminalLevelSet{\pi}:=\{\xV \in \constrainedStateSpace \, | \, \terminalCostf{\xV}{s}\leq \pi \}$ with some $\pi>0$. The results on inherent robustness only rely on the stabilizing warm-starts $\uSW{\xo}\in\admissibleWarmStartSpace{\xo}$ and the basic stability Assumptions~\ref{assum:continuity_of_transition_map}-\ref{assum:controlled_forward_invariant_terminal_set}. Therefore, suboptimal MBMPC inherits the robustness results from~\cite{Allan2017}.
\end{rem}

Depending on the parameter $M$, it is very likely that there are no better solutions than the warm-start. In such cases, the optimizer cannot improve the closed-loop performance. The next section presents an approach that allows the optimizer to start its routine at the warm-started control sequence and then improve it incrementally. 


\section{Offset input move-blocking}
\label{sec:improving_warmstart}
A straightforward approach to replicate the warm-start by optimization follows from the redefinition of the input parameterization similar to~\cite{Ong2014,Shekhar2015} with $\lambda\in\mathbb{R}_0$: 
\begin{equation}
	\uS = \unblockFunctionf{\uSBlocked,\uSW{\xo},\lambda}{b}:=\big( \matrixBar{B} \otimes \matrixBar{I}_{\inputDim} \big)\, \uSBlocked + \lambda\,\uSW{\xo}.
	\label{eq:input_paramterization_offset_mb}
\end{equation} 
In addition, the corresponding OCP needs to include the parameter $\lambda$ as another optimization parameter~\cite{Ong2014,Shekhar2015}:
\begin{equation}
	\begin{gathered}
		\min_{{\uSBlocked,\lambda}} \totalCostf{\xo}{\unblockFunctionf{\uSBlocked,\uSW{\xo},\lambda}{b}}{B}, \\ 
		\mathrm{subject~to~} \unblockFunctionf{\uSBlocked,\uSW{\xo},\lambda}{b}\in\admissibleInputSpaceSubOptimal{\xo}{N}.  
	\end{gathered}
	\label{eq:suboptimal_offset_mb_ocp}
\end{equation}
A suboptimal solution tuple is denoted by $(\uSBlockedSubOpt{\xo},\lambda^\symbolSubOpt)$. The corresponding suboptimal control sequence directly follows by: 
\begin{equation}
	\uSSubOpt{\xo} := \unblockFunctionf{\uSBlockedSubOpt{\xo},\uSW{\xo},\lambda^\symbolSubOpt}{b}.
	\label{eq:offset_level}
\end{equation}

\begin{prop}[Properties of offset MBMPC]
	Suppose Assumptions~\ref{assum:continuity_of_transition_map}-\ref{assum:initial_extended_state} hold. Assume that the suboptimal control sequences are generated based on equation~\eqref{eq:offset_level} and the suboptimal solutions to OCP~\eqref{eq:suboptimal_offset_mb_ocp}. Then, the feasible set~$\admissibleTotalSpace$ is positive invariant and the origin asymptotically stable in $\admissibleTotalSpace$ for the closed-loop system~\eqref{eq:difference_inclusion}. 
\end{prop}
\begin{pf}
 The optimizer can now resort to the warm-start $\uSW{\xo}\in\admissibleWarmStartSpace{\xo}$ autonomously by setting $\lambda^\symbolSubOpt=1$ and $\uSBlockedSubOpt{\xo} \in \{0\}^{mM}$ (see also~\cite[Thm. 1]{Shekhar2015}). By Assumption~\ref{assum:constrained_sets}, $\{0\}^{mM}\subset \constrainedInputSpace^M$. Assumption~\ref{assum:initial_extended_state} provides the first admissible warm-start at closed-loop time $n=0$. Since $\uSSubOpt{\xo}=\uSW{\xo}\in\admissibleInputSpaceSubOptimal{\xo}{N}$,  $\xmuplus\in\feasibleStateSpace$ holds after applying the suboptimal control vector $\ukfSubOpt{0}{\xo}$ for closed-loop control. From equation~\eqref{eq:generate_warmstarts} it follows that $\uSW{\xmuplus}=\totalWarmStartFunction{\xo,\uSW{\xo}}{b}\in\admissibleWarmStartSpace{\xmuplus}$ such that $\ubar{z}^+\in\admissibleTotalSpace$ (see~\cite[Proof Thm.\,14]{Allan2017}). The rest of the Proof (asymptotic stability) follows from~\cite[Prop.\,13 and Proof of Thm.\,14]{Allan2017} and relies on the Assumptions~\ref{assum:continuity_of_transition_map}-\ref{assum:controlled_forward_invariant_terminal_set}. 
\end{pf}
The main difference to~\cite{Shekhar2015} is that this subsection directly incorporates stabilizing warm-starts according to~\eqref{eq:generate_warmstarts}. Hence, the setting $\lambda^\symbolSubOpt=1$ and $\uSBlockedSubOpt{\xo} \in \{0\}^{mM}$ ensures both recursive feasibility and asymptotic stability, while the authors of~\cite{Shekhar2015} only focus on recursive feasibility. Furthermore, since the closed-loop properties rely on a suboptimal formulation, the following configuration enables a deterministic computing time without losing important closed-loop properties. Let $\lambda_i$ and $\lambda_i^+$ be the values of $\lambda$ after $i\in\mathbb{N}_0$ optimization iterations at time instance $n$ and $n+1$, respectively. Analogous relation applies for $\uSBlocked_i^+$. By manually setting $\lambda_0^+:=1$ and $\uSBlocked_0^+ \in \{0\}^{mM}$ at time instance~$n$, the optimization algorithm always starts its routine at the warm-start $\uSW{\xo}$, which is inherently an element of $\admissibleInputSpaceSubOptimal{\xo}{N}$. This warm-starting strategy of auxiliary parameters is consistent with the basic idea in suboptimal MPC according to~\cite{Pannocchia2011,Allan2017} (see~\eqref{eq:generate_warmstarts}), however, does not affect the evolution of the difference inclusion~\eqref{eq:difference_inclusion}.
\begin{rem}[Shifting blocking pattern]
	The input parameterization in~\eqref{eq:input_paramterization_offset_mb} can be extended such that it includes a time-varying blocking matrix whose evolution might follow the shifting strategy in~\cite{Cagienard2007} or in~\cite{Gonzalez2020}. Note that such shifting strategies affect the sparsity pattern of first- or second-order derivative information. A hypergraph formulation enables an efficient online sparsity exploitation and addresses the use of sparse solvers~\cite{Rosmann2018}.
\end{rem}
\begin{rem}[Regularization]
	Problem \eqref{eq:suboptimal_offset_mb_ocp} can be regularized by including $\eta\,(\lambda-1)^2$ as an additional cost term with some small weighting parameter $\eta>0$.
\end{rem}

\section{Numerical Evaluation}
\label{sec:numerical_evaluation}

\begin{figure}[t]
	\centering
	\setlength\figureheight{0.7\columnwidth} 
	\setlength\figurewidth{0.9\columnwidth} 
	\begin{tikzpicture}


\begin{axis}[%
	width=0.9\figurewidth,
	height=0.9\figureheight,
	axis y line = left,
	axis x line = bottom,
	at={(0\figurewidth,0.0\figurewidth)},
	scale only axis,
	separate axis lines,
	every outer x axis line/.append style={black},
	every x tick label/.append style={font=\color{black},font=\small},
	xmin=-1,
	xmax=1,
	y label style={at={(-0.12,0.5)}},
	every outer y axis line/.append style={draw=none},
	every y tick label/.append style={font=\color{black},font=\small},
	ymin=-1,
	ymax=0.7,
	line join=round,
	xlabel={$k$},
	ylabel={},
	ytick = {-1},
	yticklabels={},
	xtick = {-1,-0.5,0,0.5,1.0},
	xticklabels={$0$,$1$,$2$,$3$,$4$},
	axis background/.style={fill=white},
	legend columns = 0,
	legend entries={},
	legend style={at={(0.05,1.01)},anchor=south west,legend cell align=left,align=left,draw=none,font=\footnotesize,fill opacity=0.85}
	]

\addplot [color=black,densely dashed,line width=1.0pt] coordinates {(-1.0,-0.9) (-0.5,-0.9) (0.0,-0.9) (0.0,-0.5) (1.0,-0.5)};
\addplot [thick,color=black,fill=gray3, line width=0.6pt,only marks,mark=*, mark size=2pt] coordinates {(-1.0,-0.9)};
\addplot [thick,color=black,fill=white, line width=0.6pt,only marks,mark=*, mark size=2pt] coordinates {(-0.5,-0.9)};
\addplot [thick,color=black,fill=gray3, line width=0.6pt,only marks,mark=*, mark size=2pt] coordinates {(0.0,-0.5)};
\addplot [thick,color=black,fill=white, line width=0.6pt,only marks,mark=*, mark size=2pt] coordinates {(0.5,-0.5)};
\node (source) at(-0.85,-0.8){$\ukBlocked{0}$};
\node (source) at(0.15,-0.4){$\ukBlocked{1}$};

\addplot [color=black,densely dashed,line width=1.0pt] coordinates {(-1.0,-0.5) (-0.5,-0.5) (-0.5,-0.1) (0.5,-0.1) (0.5,-0.25) (1.0,-0.25)};
\addplot [thick,color=black,fill=gray3, line width=0.6pt,only marks,mark=*, mark size=2pt] coordinates {(-1.0,-0.5)};
\addplot [thick,color=black,fill=gray3, line width=0.6pt,only marks,mark=*, mark size=2pt] coordinates {(-0.5,-0.1)};
\addplot [thick,color=black,fill=gray3, line width=0.6pt,only marks,mark=*, mark size=2pt] coordinates {(0.0,-0.1)};
\addplot [thick,color=black,fill=gray3, line width=0.6pt,only marks,mark=*, mark size=2pt] coordinates {(0.5,-0.25)};
\node (source) at(-0.8,-0.4){$\ukfW{0}{\xo}$};
\node (source) at(-0.3,0.0){$\ukfW{1}{\xo}$};
\node (source) at(0.2,0.0){$\ukfW{2}{\xo}$};
\node (source) at(0.7,-0.15){$\ukfW{3}{\xo}$};

\node (source) at(-0.15,0.65){\scriptsize $\ukBlocked{0} +\lambda \max\{\ukfW{0}{\xo},\ukfW{1}{\xo}\}\leq \max\constrainedInputSpace$};
\node (source) at(-0.165,0.5){\scriptsize $\ukBlocked{0} +\lambda \min\{\ukfW{0}{\xo},\ukfW{1}{\xo}\}\geq \min\constrainedInputSpace$};
\node (source) at(-0.15,0.35){\scriptsize $\ukBlocked{1} +\lambda \max\{\ukfW{2}{\xo},\ukfW{3}{\xo}\}\leq \max\constrainedInputSpace$};
\node (source) at(-0.165,0.2){\scriptsize $\ukBlocked{1} +\lambda \min\{\ukfW{2}{\xo},\ukfW{3}{\xo}\}\geq \min\constrainedInputSpace$};

\node (source) at(0,-0.3){$+$};
\end{axis}

\end{tikzpicture}%
	\caption{Example of linear inequalities in the case of offset MBMPC with $M=2$, uniform blocking, $N=4$, and $\inputDim=1$.}
	\label{fig:offset}
\end{figure}
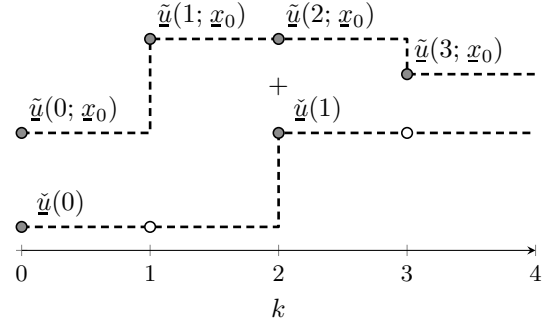

\begin{figure}[t]
	\centering
	\setlength\figureheight{0.8\columnwidth} 
	\setlength\figurewidth{0.9\columnwidth} 
	\begin{tikzpicture}


\begin{axis}[%
	width=0.9\figurewidth,
	height=0.9\figureheight,
	axis y line = left,
	axis x line = bottom,
	at={(0\figurewidth,0.0\figurewidth)},
	scale only axis,
	separate axis lines,
	every outer x axis line/.append style={black},
	every x tick label/.append style={font=\color{black},font=\small},
	xmin=-1,
	xmax=1,
	xmajorgrids,
	y label style={at={(-0.12,0.5)}},
	every outer y axis line/.append style={black},
	every y tick label/.append style={font=\color{black},font=\small},
	ymin=-1,
	ymax=1,
	line join=round,
	xlabel={\small $x_1(k),\,x_{\mu,1}(n)$},
	ylabel={\small $x_2(k),\,x_{\mu,2}(n)$},
	ymajorgrids,
	axis background/.style={fill=white},
	legend columns = 5,
	legend entries={ T0, T1, T2, T3, T4, T5, T6},
	legend style={at={(0.05,1.01)},anchor=south west,legend cell align=left,align=left,draw=none,font=\footnotesize,fill opacity=0.85}
	]

\addlegendimage{color=black,densely dashed,line width=1.2pt};	
\addlegendimage{color=black,solid,line width=1.2pt};
\addlegendimage{color=gray5,solid,line width=1.2pt};
\addlegendimage{color=gray1,solid,line width=1.2pt};
\addlegendimage{color=black,densely dotted,line width=1.2pt};
\addlegendimage{color=gray4,densely dashdotted,line width=1.2pt};
\addlegendimage{color=gray1,densely dotted,line width=1.2pt};

\addplot[color=gray4,solid,line width=0.8pt]
table[]{mpc_feasible_polytope.txt};

\addplot[color=gray2,solid,line width=0.8pt]
table[]{mb_mpc_feasible_polytope.txt};

\addplot[color=gray2,solid,line width=0.8pt]
table[]{terminal_ball.txt};

\addplot [thick,color=black,fill=black, line width=0.6pt,only marks,mark=x, mark size=2pt] coordinates {(0.8,0.0)};

\addplot[color=gray1,densely dotted,line width=1.0pt]
table[]{offset_mb_m16_x0_1_iter_3_state_space_closed_loop.txt};

\addplot[color=gray4,densely dashdotted,line width=1.0pt]
table[]{mpc_x0_1_state_space_closed_loop.txt};

\addplot[color=gray1,solid,line width=1.8pt]
table[]{offset_mb_x0_1_iter_0_state_space_closed_loop.txt};

\addplot[color=gray5,solid,line width=0.8pt]
table[]{mb_x0_1_buffer_state_space_closed_loop.txt};

\addplot[color=black,densely dotted ,line width=1.0pt]
table[]{offset_mb_x0_1_iter_3_state_space_closed_loop.txt};

\addplot[color=black,densely dashed,line width=1.0pt]
table[]{mb_x0_1_open_loop_run_no_1.txt};

\addplot[color=black,solid,line width=0.8pt]
table[]{mb_x0_1_state_space_closed_loop.txt};

\node (source) at(0.0,0.6){\scriptsize \textcolor{black}{$\bullet\approx\feasibleStateSpaceMB$}};
\node (source) at(0.01,0.45){\scriptsize \textcolor{black}{$(M=2)$}};

\node (source) at(-0.03,0.95){\scriptsize \textcolor{black}{$\bullet=\constrainedStateSpace$}};

\node (source) at(0.6,-0.95){\scriptsize \textcolor{black}{$\approx\feasibleStateSpace$}};
\node (destination) at(0.35,-0.8){};
\draw[->](source)--(destination);

\node (source) at(0.17,0.20){\scriptsize \textcolor{gray3}{$\terminalLevelSet{\pi}$}};

\end{axis}


\begin{axis}[%
	width=0.25\figurewidth,
	height=0.25\figureheight,
	/pgf/number format/use comma,
	axis y line = left,
	axis x line = bottom,
	at={(0.60\figurewidth,0.55\figurewidth)},
	scale only axis,
	separate axis lines,
	every outer x axis line/.append style={black},
	every x tick label/.append style={font=\color{black},font=\tiny},
	xmin=1,
	xmax=200,
	xmajorgrids,
	xtick={1,200},
	xticklabels={0,200},
	yticklabels={30,0,},
	ytick={30,0,-1},
	y label style={at={(-0.05,0.5)}},
	x label style={at={(0.5,-0.05)}},
	every outer y axis line/.append style={black},
	every y tick label/.append style={font=\color{black},font=\tiny},
	ymin=-1,
	ymax=30,
	ylabel={\scriptsize{Costs}},
	xlabel={\scriptsize{$n$}},
	line join=round,
	ymajorgrids,
	axis background/.style={fill=white,fill opacity=0.85},
	legend columns = 1,
	legend entries={},
	legend style={at={(0.75,1)},anchor=south west,legend cell align=left,align=left,draw=none,font=\scriptsize}
	]
	
	\addplot[color=gray1,densely dotted,line width=1.0pt]
	table[]{offset_mb_m16__x0_1_iter_3_costs.txt};
	
	\addplot[color=gray4,densely dashdotted,line width=1.0pt]
	table[]{mpc_x0_1_costs.txt};
	
	\addplot[color=gray1,solid,line width=1.8pt]
	table[]{offset_mb_x0_1_iter_0_costs.txt};
	
	\addplot[color=gray5,solid,line width=0.8pt]
	table[]{mb_x0_1_buffer_costs.txt};
	
	\addplot[color=black,solid,line width=0.8pt]
	table[]{mb_x0_1_costs.txt};
	
	\addplot[color=black,densely dotted,line width=1.0pt]
	table[]{offset_mb_x0_1_iter_3_costs.txt};

\end{axis}


\begin{axis}[%
	width=0.25\figurewidth,
	height=0.25\figureheight,
	axis y line = left,
	axis x line = bottom,
	at={(0.11\figurewidth,0.55\figurewidth)},
	scale only axis,
	separate axis lines,
	every outer x axis line/.append style={black},
	every x tick label/.append style={font=\color{black},font=\tiny},
	xmin=-0.05,
	xmax=0.05,
	xmajorgrids,
	xtick={-0.05,0,0.05},
	xticklabels={-0.05,,0.05},
	yticklabels={-0.05,,0.05},
	ytick={-0.05,0,0.05},
	y label style={at={(-0.05,0.5)}},
	x label style={at={(0.5,-0.05)}},
	every outer y axis line/.append style={black},
	every y tick label/.append style={font=\color{black},font=\tiny},
	ymin=-0.05,
	ymax=0.05,
	ylabel={\scriptsize{$x_{\mu,2}(n)$}},
	xlabel={\scriptsize{$x_{\mu,1}(n)$}},
	line join=round,
	scaled ticks=false,
	ymajorgrids,
	axis background/.style={fill=white,fill opacity=0.85},
	legend columns = 1,
	legend entries={},
	legend style={at={(0.75,1)},anchor=south west,legend cell align=left,align=left,draw=none,font=\scriptsize}
	]
	
	\addplot[color=gray2,solid,line width=0.8pt]
	table[]{terminal_ball.txt};
	
	\addplot[color=gray1,densely dotted,line width=1.0pt]
	table[]{offset_mb_m16_x0_1_iter_3_state_space_closed_loop.txt};
	
	\addplot[color=gray4,densely dashdotted,line width=1.0pt]
	table[]{mpc_x0_1_state_space_closed_loop.txt};
	
	\addplot[color=gray1,solid,line width=1.8pt]
	table[]{offset_mb_x0_1_iter_0_state_space_closed_loop.txt};
	
	\addplot[color=gray5,solid,line width=0.8pt]
	table[]{mb_x0_1_buffer_state_space_closed_loop.txt};
	
	\addplot[color=black,densely dotted,line width=1.0pt]
	table[]{offset_mb_x0_1_iter_3_state_space_closed_loop.txt};
	
	\addplot[color=black,densely dashed,line width=0.8pt]
	table[]{mb_x0_1_open_loop_run_no_1.txt};
	
	\addplot[color=black,solid,line width=1.0pt]
	table[]{mb_x0_1_state_space_closed_loop.txt};
	
\end{axis}


\begin{axis}[%
	width=0.25\figurewidth,
	height=0.25\figureheight,
	/pgf/number format/use comma,
	axis y line = left,
	axis x line = bottom,
	at={(0.11\figurewidth,0.06\figurewidth)},
	scale only axis,
	separate axis lines,
	every outer x axis line/.append style={black},
	every x tick label/.append style={font=\color{black},font=\tiny},
	xmin=1,
	xmax=150,
	xmajorgrids,
	xtick={1,150},
	xticklabels={0,150},
	yticklabels={,1,0.6},
	ytick={1.05,1,0.6},
	y label style={at={(-0.05,0.5)}},
	x label style={at={(0.5,-0.05)}},
	every outer y axis line/.append style={black},
	every y tick label/.append style={font=\color{black},font=\tiny},
	ymin=0.6,
	ymax=1.05,
	ylabel={\scriptsize{$\lambda_3$}},
	xlabel={\scriptsize{$n$}},
	line join=round,
	ymajorgrids,
	axis background/.style={fill=white,fill opacity=0.85},
	legend columns = 1,
	legend entries={},
	legend style={at={(0.75,1)},anchor=south west,legend cell align=left,align=left,draw=none,font=\scriptsize}
	]
	
	\addplot[color=gray1,densely dotted,line width=0.8pt]
	table[]{lambda_i3_m16.txt};
	
	\addplot[color=black,densely dotted,line width=0.8pt]
	table[]{lambda_i3_m2.txt};
	
\end{axis}

\end{tikzpicture}%
	\caption{Phase portrait for the benchmark system~\eqref{eq:benchmark_system}.}
	\label{fig:phaseLevel}
\end{figure}
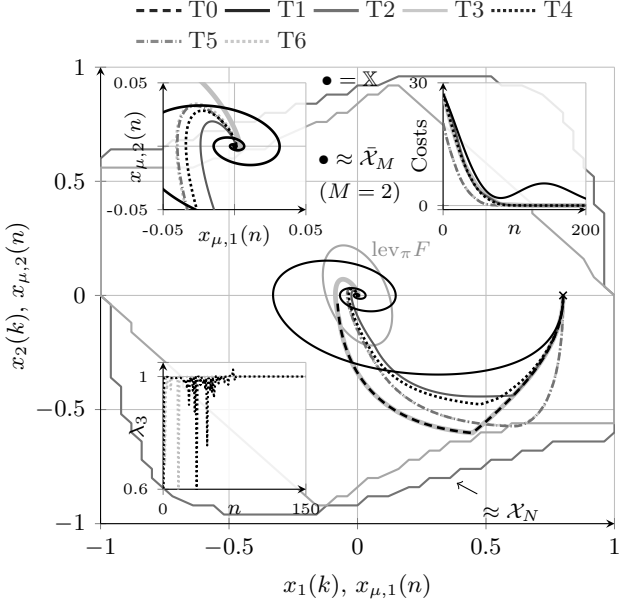

The numerical evaluation is based on a discrete-time version of the Van der Pol oscillator, resulting from the application of Euler's method with a step size of $t_\mathrm{s}=2^{-5}\,\mathrm{s}$: 
\begin{align}
	x_1(k+1) &=  x_1(k) + t_\mathrm{s} x_2(k), \nonumber \\
	x_2(k+1) &=  x_2(k) + t_\mathrm{s} u(k) - t_\mathrm{s} x_1(k) + \label{eq:benchmark_system}\\
	& \quad+t_\mathrm{s}x_2(k)\big(1-x_1^2(k)\big).\nonumber 
\end{align}
Here, $\xk{k}=\big(x_1(k), x_2(k) \big)^\intercal$ and $u(k)\in\mathbb{R}^1$. The cost functions are defined by $\stageCostf{\xk{k}}{\uk{k}}{b}:=\normEQ{\xk{k}}{Q} + 0.1\,u^2(k)$ with $\matrixBar{Q}=\mathrm{diag}(1,0.1)$ and $\terminalCostf{\xk{N}}{b}:=\normEQ{\xk{N}}{P}$. The matrix $\matrixBar{P}$ is the solution to the discrete Riccati equation $\matrixBar{P}=\matrixBar{A}^\intercal\matrixBar{P}\matrixBar{A}-(\matrixBar{A}^\intercal\matrixBar{P}\matrixBar{B})(\rho\matrixBar{R}+\matrixBar{B}^\intercal\matrixBar{P}\matrixBar{B})^{-1}(\matrixBar{B}^\intercal\matrixBar{P}\matrixBar{A})+\rho\matrixBar{Q}$. The pair $(\matrixBar{A},\matrixBar{B})$ is controllable and thus stabilizable with $\matrixBar{A}=\big(\partial\fv/\partial\xk{\cdot}\big)\refpairzero$ and $\matrixBar{B}=\big(\partial\fv/\partial\uk{\cdot}\big)\refpairzero$. The local control law is defined by $\localControlLawf{\xk{k}}:=-\matrixBar{K}\xk{k}$ with $\matrixBar{K}=(\rho\matrixBar{R}+\matrixBar{B}^\intercal\matrixBar{P}\matrixBar{B})^{-1}\matrixBar{B}^\intercal\matrixBar{P}\matrixBar{A}$. Without a proof, the terminal set is defined by $\terminalSet:=\terminalLevelSet{\pi}$ with $\pi=0.4856$ and $\rho = 1.001$ (see,\,e.g,\,\cite[Sec.\,2.5.5]{Rawlings2020}). This section only applies uniform input move-blocking for a horizon of $N=80$. Multiple shooting transforms the corresponding discrete-time OCPs into nonlinear programs \cite{Bock1984}. There exist as many shooting intervals $S=M$ as there are blocking intervals. Here, shooting and blocking intervals have the same length. An appropriate generic software framework (written in MATLAB\textsuperscript{\tiny\textregistered}) exploits the sparsity of the Jacobian. The framework interfaces with the general purpose solver IPOPT~\cite{Waechter2006} and relies on its Hessian approximation. Finite central differences are used to approximate function derivatives. The additional optimization parameter $\lambda$ leads to a dense column vector in the Jacobian of the constraints. In addition, linear inequalities are required to ensure that $\unblockFunctionf{\uSBlocked,\uSW{\xo},\lambda}{b}\in \constrainedInputSpace^N$. Figure~\ref{fig:offset} visualizes the implication of the linear inequalities for input box-constraints. Constraints are defined as $-1\leq x_1(k) \leq 1$, $-1\leq x_2(k) \leq 1$, and $|u(k)|\leq 1$. 
Figure~\ref{fig:phaseLevel} shows the phase portrait for system~\eqref{eq:benchmark_system}, including the first open-loop solution (T0) with $M=2$ and different closed-loop state trajectories. Although the standard MBMPC (T1), which does not rely on stabilizing warm starts, shows recursive feasibility in this particular numerical experiment, the corresponding cost function $\totalCost$ does not decrease in the sense of Lyapunov. The first approach from Section~\ref{sec:suboptimal_mpc} (T2) applies the fallback solutions stored in the buffer at most time instances~$n$. However, since the closed-loop trajectory does not match the first prediction, the optimizer also finds solutions better than the fallback solutions. Offset move-blocking from Section~\ref{sec:improving_warmstart} is illustrated for two configurations. In the first setting (T3), offset MBMPC simply reuses the stabilizing warm-start since the maximal number of optimization iterations is set to $i=0$. In the second case, the maximal number of optimization iterations is increased to $i=3$ (T4). Figure~\ref{fig:phaseLevel} demonstrates that the optimizer improves the warm-start since the open-loop and closed-loop solutions no longer coincide and the subplot shows changing values for~$\lambda_3$. For both proposed approaches, the cost function $\totalCost$ decreases in the sense of Lyapunov. Because of the very restrictive input parameterization of the previous trajectories with $M=2$, common MPC (T5) with $M=N=80$ shows a superior closed-loop control performance. However, with the setting $M=16$ and $i=3$ offset MBMPC reaches nearly the same control performance (T6). 
The statistical evaluation of computing times considers 100 cold-started open-loop optimization runs at $n=0$. Let $t_\mathrm{m}$ and $t_\mathrm{q}$ be the median and the 0.95-quantile, respectively, of the measured time values with MPC. For the presented numerical setup, MPC reaches its fastest performance with $S=M=80$ (full discretization). MBMPC with $S=M=2$ results in $0.3\,t_\mathrm{m}$ and $0.3\,t_\mathrm{q}$. The overhead for evaluating the buffer is negligible. Offset MBMPC requires an overhead for the additional linear inequalities (see Fig.~\ref{fig:offset}) and the parameter $\lambda$ and results in $0.57\,t_\mathrm{m}$ and $0.58\,t_\mathrm{q}$. The moderate reduction to $M=16$ addresses practical applications and results in $0.78\,t_\mathrm{m}$ and $0.82\,t_\mathrm{q}$.

\section{Conclusion}
\label{sec:conclusion}

This paper first provides a literature review on online MBMPC with a receding horizon. Since asymptotic stability with MBMPC for discrete-time systems is still an open problem, this contribution transfers the results of suboptimal MPC to the formulation of MBMPC. By explicitly classifying the blocked input parameterization as a source of suboptimality, recursive feasibility and asymptotic stability of the origin directly follow from stabilizing warm-starts according to~\cite{Pannocchia2011,Allan2017,Rawlings2020}. The strategy for embedding the solution of the previous closed-loop time instance according to~\cite{Ong2014,Shekhar2015} eliminates the need to hold back a stabilizing fallback solution. The proposed approach makes it possible to easily reduce the computing time without losing the benefits of stabilizing terminal conditions. The blocking pattern and the number of degrees of freedom in control can be chosen depending on the application.  

\bibliographystyle{ieeetr}        
\bibliography{ms}                 

\end{document}